Perceptions of Discriminatory Decisions of Artificial Intelligence: Unpacking the Role of

Individual Characteristics

Soojong Kim[1]

[1] Department of Communication, University of California Davis

**Author Note**

Soojong Kim  https://orcid.org/0000-0002-1334-5310

Correspondence concerning this article should be addressed to Soojong Kim, University

of California Davis, 361 Kerr Hall, Davis, CA 95616, United States. Email: sjokim@ucdavis.edu

**Abstract**

This study investigates how personal differences (digital self-efficacy, technical knowledge, belief in equality, political ideology) and demographic factors (age, education, and income) are associated with perceptions of artificial intelligence (AI) outcomes exhibiting gender and racial bias and with general attitudes towards AI. Analyses of a large-scale experiment dataset ($N =$ 1,206) indicate that digital self-efficacy and technical knowledge are positively associated with attitudes toward AI, while liberal ideologies are negatively associated with outcome trust, higher negative emotion, and greater skepticism. Furthermore, age and income are closely connected to cognitive gaps in understanding discriminatory AI outcomes. These findings highlight the importance of promoting digital literacy skills and enhancing digital self-efficacy to maintain trust in AI and beliefs in AI usefulness and safety. The findings also suggest that the disparities in understanding problematic AI outcomes may be aligned with economic inequalities and generational gaps in society. Overall, this study sheds light on the socio-technological system in which complex interactions occur between social hierarchies, divisions, and machines that reflect and exacerbate the disparities.

Perceptions of Discriminatory Decisions of Artificial Intelligence: Unpacking the Role of

Individual Characteristics

The use of artificial intelligence (AI) and automated decision-making has surged in recent years, and these technologies are replacing human decision-making processes in various domains, including the news media, healthcare, law, finance, and law enforcement (e.g., Diakopoulos & Koliska, 2017; Jha & Topol, 2016; Kennedy et al., 2011; Nissan, 2017; Thurman et al., 2019; Yu & Kohane, 2019). While the socio-technological shift has the potential to bring about positive changes, such as increased efficiency, productivity, and innovation at workplaces and in daily lives, concerns are also mounting regarding discriminatory and unjust outcomes that the new technologies could produce (e.g., Asplund et al., 2020; Imana et al., 2021).

Despite the potential benefits, user interactions with AI and algorithms often reveal notable challenges, such as unfairness, distrust, skepticism, and negative emotions experienced by users. Research in Human-Computer Interaction (HCI) has demonstrated that these challenges are not only technical but also rooted in user engagement and the contexts of use. Especially, a major body of work in HCI closely examines user interactions with algorithms and AI systems, focusing on the complexities of human-algorithm interaction (Choung, Seberger, et al., 2023; Hajigholam Saryazdi, 2024; Lee & Baykal, 2017; Ochmann et al., 2024; Rader & Gray, 2015; Starke et al., 2022; Wang et al., 2020). For instance, Binns et al. (2018) discuss justice perceptions in algorithmic decisions, highlighting how the lack of transparency and the reduction of human qualities to mere statistical percentages can exacerbate feelings of unfairness and indignity among users. Choung et al. (2023) explore how perceptions of AI in decision-making can vary based on the context and outcome, finding that AI decisions are often viewed as more trustworthy, fair, and useful than human decisions especially when rejecting candidates in a job application scenario. This domain of HCI research stresses the need to understand challenges in enhancing interactions with AI systems, identify factors influencing the interactions, and develop and implement human-centered, culturally aware design solutions that address these challenges.

It has been known that AI and algorithms have the possibility of producing unequal, unfair, and even unjust outcomes (e.g., Barocas et al., 2019; Hooker, 2021), which could perpetuate and even exacerbate social inequalities and injustices (Benjamin, 2019; Noble, 2018; O'Neil, 2016). Systematic problems that exist in society can also easily penetrate and be ingrained into AI and algorithms (e.g., Koenecke et al., 2020; Obermeyer et al., 2019). Examples of algorithmic racial and gender discrimination have been observed in various domains, including housing markets (Asplund et al., 2020; J. Miller, 2020), job markets (Chen et al., 2018; Imana et al., 2021), human recognition (Koenecke et al., 2020; Menezes et al., 2021), and web search engines (Makhortykh et al., 2021; Vlasceanu & Amodio, 2022). These alarming instances illustrate the imperative to assess the social consequences and implications of machine-driven discrimination and injustice, to understand public attitudes and perceptions toward algorithms, AI technologies, and public policy governing their use, and to improve the design and implementation for more equitable and just socio-technological systems (Angerschmid et al., 2022; Dietvorst et al., 2014; Dolata et al., 2022; Logg et al., 2019).

Focusing on the imperative to understand public attitudes and perceptions toward algorithms and AI, the present study investigates factors that shape perceptions of discriminatory AI outcomes, and general attitudes toward AI in the contexts of machine-augmented discrimination. Previous research has investigated how individuals' personal differences (e.g., self-efficacy, knowledge, political beliefs) and demographic factors (e.g., education, age, and income) are associated with their attitudes toward AI use, such as trust in AI, perceived risk of AI use, and the preference of AI over human decision-making (Araujo et al., 2020; Lee & Rich, 2021; Logg et al., 2019; Thurman et al., 2019). However, still little is known about how individual differences influence perceptions of discriminatory outcomes produced by algorithms and AI, although unfair and biased outcomes from automated systems are increasingly regarded as prevalent and harmful (J. Miller, 2020; Thune, 2022; Verma, 2022). The role of demographic factors also suffers a lack of investigation in the domain of the public perception of algorithms and AI (Starke et al., 2022). To address these gaps of knowledge and to provide a more comprehensive understanding of the factors connected with perceptions of discrimination generated or enhanced by machines, we analyzed data obtained from a large-scale experiment ($N$ = 1,206) that examined perceptions of gender and racial biases in AI systems. This study focused on how personal differences and demographic factors are associated with cognitive and emotional responses to automated decisions that discriminate against certain gender or racial groups.

The remainder of this paper is structured as follows. We first review previous research on the perceptions of algorithms and AI and articulate the research hypotheses and questions of the study. The methodology and results of the research are then presented. Lastly, the discussion of the findings, the limitations of the research, and potential directions for future research follow.

## Public Perceptions of AI

The current research explores the dynamics of user interactions with AI systems, which is one of the central goals of HCI research. Especially, it investigates specific difficulties that users face in using AI systems—the users' perception of unfairness, distrust, skepticism, and negative emotion toward AI outcomes—which also influence their interaction with the systems. By doing so, this study contributes to the important literature on how users' perceptions of algorithms and algorithmic outcomes are shaped and can be improved (Angerschmid et al., 2022; Araujo et al., 2020; Binns et al., 2018; Dietvorst et al., 2014; Lee & Baykal, 2017; Rader & Gray, 2015; Wang et al., 2020). The present research specifically focuses on the perceptions of discriminatory outcomes and examines how the perceptions of AI outcomes are influenced by individual characteristics, aiming to propose targeted design remedies based on these insights.

Previous research showed that people often refuse to use algorithms and AI, despite generally positive attitudes toward them (Logg et al., 2019; Thurman et al., 2019). For example, one study found an aversion to AI-based medical decisions compared with decisions made by human healthcare providers (Longoni et al., 2019). Errors in algorithms and their potential significance can contribute to the aversion and refusal to adopt algorithms: Observing errors can reduce confidence and intention to choose algorithms over human decision-makers, even in situations where algorithms outperform humans (Dietvorst et al., 2014). Concerns about potential biases may also harm perceived fairness and trust in these systems (Araujo et al., 2020; Lee & Baykal, 2017). Several cognitive and emotional dimensions have been studied as cognitive indicators of algorithm appreciation and aversion, including perceived fairness, trust in algorithms and AI, and emotional reactions (Angerschmid et al., 2022; Araujo et al., 2020; Choung, David, et al., 2023; Lee, 2018; Lee & Baykal, 2017; Lee & Rich, 2021; Wang et al., 2020).

**Perceptions of Discriminatory AI Outcomes**

Previous studies have explored how individual characteristics, including personal differences (such as technological self-efficacy, domain-specific knowledge, and political beliefs) and demographic factors (such as race, gender, education, and age), shape people's attitudes toward algorithms and AI (Araujo et al., 2020; Lee & Rich, 2021; Logg et al., 2019;

Thurman et al., 2019) and how these factors affect perceptions of automated outcomes (Lee, 2018). However, little research has focused on the role of individual characteristics in cases where people encounter discriminatory algorithm outcomes. This is a critical gap in our understanding, given the importance of understanding public perceptions of biased machine decisions for developing policies and technologies that are more inclusive and beneficial for society (Benjamin, 2019; Noble, 2018; O'Neil, 2016). Examining perceptions of algorithmic discrimination can also help advance long-standing efforts in social sciences to answer essential questions on social identities, attribution processes, and cognitive biases in this "age of AI" (Kissinger et al., 2021). Hence, the present research aimed to gain a more detailed and in-depth understanding of the public perception of algorithms and AI, focusing on four aspects of individual differences (digital self-efficacy, technical knowledge, belief in equality, and political ideology) and three demographic factors (education, age, and income.)

*Digital Self-Efficacy*

Self-efficacy has been established as a crucial predictor of attitudes toward certain technology and its use (Aesaert & van Braak, 2014; Rimal, 2001). A study suggested that online self-efficacy, an individual's belief in their own ability to protect personal information while using the Internet, is linked to perceived fairness, usefulness, and risk of algorithms (Araujo et al., 2020). The current research explores the role of digital self-efficacy by expanding the concept of online self-efficacy. Digital self-efficacy is defined as self-efficacy in safeguarding personal information when using digital devices and services. General concerns about data collection and the use of digital technologies are known to be associated with attitudes toward algorithms (Thurman et al., 2019). With algorithms and AI increasingly integrated into all types and modes of digital technologies, digital self-efficacy is expected to encompass self-efficacy not only during web browsing and general Internet use, but also in the rapidly evolving technological landscape.

This research examined if perceptions of discriminatory AI outcomes are associated with digital self-efficacy. Based on the prior work, it was assumed that people with higher digital self-efficacy interpret discriminatory AI outcomes less negatively. This is because people's higher confidence in their ability to understand how technologies utilize personal information and how to protect themselves from potential negativities may encourage them to perceive a biased outcome less negatively, viewing it as less unfair and more trustworthy with less negative emotion and skepticism.

H1: Digital self-efficacy is positively associated with perceived fairness and trust of discriminatory AI outcomes and negatively associated with negative emotion and skepticism toward the outcomes.

*Technical Knowledge*

Domain-specific knowledge of technology has been linked to general attitudes toward technologies, but previous studies have reported mixed findings on the direction of this relationship. Some studies have identified a positive relationship between technical knowledge and perceptions of the usefulness and fairness of AI (Araujo et al., 2020). Comfort with mathematics, which could be closely related to domain-specific knowledge of digital technologies and AI, has also been associated with more favorable attitudes toward algorithmic recommendations (Logg et al., 2019; Thurman et al., 2019). However, other studies have reported a decrease in perceived fairness of algorithm decisions as knowledge of computer programming increased (Lee & Baykal, 2017).

The present study aims not only to add further evidence of the relationship between technical knowledge and perceptions of the technology but also to explore how this knowledge influences perceptions of discriminatory outcomes generated by the technology, an issue that has not been touched by prior investigations. Since technical knowledge is connected to an individual's ability to understand the capabilities and limitations of algorithms and AI, greater technical knowledge is associated with reduced perceived risks and increased confidence in accuracy and reliability (Said et al., 2022). Hence, individuals with greater technical knowledge, when compared with those with lower levels of technical knowledge, may view discriminatory AI outcomes as less unfair, more trustworthy, and less likely to exhibit negative emotions toward automated decisions. Hence, this study proposes the following hypothesis:

H2: Technical knowledge is positively associated with perceived fairness and trust of discriminatory AI outcomes and negatively associated with negative emotion and skepticism toward them.

*Belief in Equality*

Belief in equality has been found to affect individuals' general attitudes toward algorithms. Araujo et al. (2020) showed that people with stronger beliefs in equality viewed algorithmic decision-making as fairer and more useful. When outcomes are biased, however, the nature of the association between equality belief and outcome perceptions may differ from that of general attitudes toward AI. This is because the violation of equality could challenge the common heuristic that machines are unbiased and objective (Sundar, 2008), and biased machine decisions could also make people doubtful about decision-making procedures and agencies (Lee, 2018; Skarlicki & Folger, 1997).

The attribution theory helps further understand potential associations between equality belief and perceptions of discriminatory AI outcomes. People are more inclined to assign responsibility to a causal agent when the outcome is perceived as more severe (Robbennolt, 2000; Schroeder & Linder, 1976; Walster, 1966). Reactions to intentional discriminatory acts are more likely to be negative than those toward unintentional ones (Stouten et al., 2006; Swim et al., 2003). Therefore, individuals with stronger beliefs in equality, who are likely to view discriminatory situations as more problematic, may assign blame to harmful intentions embedded within automated systems rather than considering them as technical errors occurring by chance. This, in turn, may lead to more negative reactions toward biased AI outcomes. Thus, we propose the following hypothesis:

H3: Belief in equality is negatively associated with perceived fairness and trust of discriminatory AI outcomes and positively associated with negative emotion and skepticism toward them.

*Political Ideology*

Political ideology is known to relate to people's attitudes toward algorithms. Some scholars have shown that individuals with more liberal ideologies are likely to have more positive attitudes toward AI and its use, while those with more conservative ideologies tend to be more skeptical about them (Castelo & Ward, 2021; Mack et al., 2021; Peng, 2020; Zhang & Dafoe, 2019). For example, Castelo and Ward (2021) showed that conservative political beliefs were associated with low levels of comfort with and trust in AI.

Considering the unprecedented transformative power of AI on established societal structures and hierarchies, political ideology may affect individuals' cognitive and affective reactions to discriminatory AI outcomes. Unlike general attitudes toward algorithms and AI, perceptions of discriminatory outcomes may display different relationships with political ideology. Specifically, individuals espousing more liberal beliefs, which place emphasis on equality and due process (Jost et al., 2009), may perceive discriminatory outcomes as more unfair or untrustworthy than those with more conservative ideologies and hence exhibit more negative reactions to the automated results. Thus, this study posited the following hypothesis:

H4: More liberal political ideology is associated with more negative cognitive and emotional responses to discriminatory AI outcomes.

Observing bias and errors affects the appreciation, adoption, and approval of algorithms, AI, and relevant policies (Choung, David, et al., 2023; Dietvorst et al., 2014; Schiff et al., 2022). Especially, it is in discriminatory situations where general attitudes toward AI, such as perceived risk of AI, become particularly relevant. However, there is a lack of evidence regarding how personal differences relate to general AI attitudes in contexts where individuals encounter discriminatory automated outcomes. Thus, the present study explores the following research question:

RQ1: How are digital self-efficacy, technical knowledge, belief in equality, and political ideology associated with general attitudes toward AI, when experiencing discriminatory AI outcomes?

**The Role of Demographic Factors**

Different social groups may give more importance to specific aspects of AI than others (e.g., Kieslich et al., 2022). Prior work has investigated the relationship between demographic characteristics and general attitudes toward AI. For example, Helberger et al. (2020) asked participants to compare AI and humans in terms of their ability to make a fairer decision. They found that AI was evaluated higher in fairness among younger and more educated participants.

Despite recent progress, the role of demographic factors, such as age, education, and income, has not received sufficient attention and needs further exploration (Starke et al., 2022). Witnessing discrimination and errors can impact people's approval, acceptance, and appreciation of algorithms and AI, as well as their support for relevant public policies (Choung, David, et al., 2023; Dietvorst et al., 2014; Schiff et al., 2022). Situations involving discriminatory outcomes may divert and intensify individuals' attitudes and perceptions. By examining the influence of these demographic factors, we can gain deeper insights into the intricate links between economic disparities, intergenerational disparities, and cognitive disparities in individuals' perceptions of AI outcomes and attitudes toward AI. Therefore, this study investigates the following research questions regarding how demographic factors are associated with general attitudes and outcome perceptions:

RQ2: How are education, age, and income associated with general attitudes toward AI, in contexts involving discriminatory outcomes?

RQ3: How are education, age, and income associated with cognitive and emotional responses to discriminatory AI outcomes?

The present research also examined whether the perceived pervasiveness of discriminatory AI outcomes varies based on personal differences. Previous research has shown that personal differences are connected with the perception of discrimination and its pervasiveness (Bagci et al., 2017; Gong et al., 2017). The perceived pervasiveness of discrimination induced or amplified by machines could have far-reaching consequences, including affecting attitudes toward algorithms and AI, support for related policies, and various aspects of people's lives (e.g., Schmitt et al., 2003; Stroebe et al., 2011). Thus, this study explored the following research question:

RQ4: How is the perceived pervasiveness of discriminatory AI outcomes associated with digital self-efficacy, technical knowledge, belief in equality, and political ideology?

## Method

The present research sought to capture the multidimensionality of factors that may influence individuals' perceptions of discriminatory AI outcomes and general attitudes toward AI. Incorporating multiple individual characteristics, including personal differences and demographic factors, that may influence AI perceptions could be also beneficial for the analysis and the interpretation of findings. By controlling for other characteristics in examining the association between a certain individual characteristic and a variable about AI perception, one can reduce some of the confounding influences that could complicate the interpretation of the association.

### Participants

Participants were drawn from a convenient sample obtained via Prolific, an online participant recruitment platform. To be eligible for the experiment, participants had to be at least 18 years old and located in the United States. Those who met these requirements were directed to an experiment webpage. Participants who had issues with following the experimental instructions or providing the correct answer to an attention check question were excluded from the analysis. The results reported in this study were based on the remaining 1,206 participants.

Of the 1,206 participants, 73.88% ($n = 891$) self-identified as white, and 7.38% ($n = 89$) as Black. The average age was 37.90 years ($SD = 14.48$). Participants who self-identified as men, women, and non-binary were 42.95% ($n = 518$), 53.40% ($n = 644$), and 2.99% ($n = 36$), respectively. A majority (86.90%) of participants held degrees higher than high school graduates, with 17.00% having postgraduate education. The percentage of participants with an income of less than $50,000 was 45.52%.

**Procedure**

This research was based on the secondary analysis of the data obtained from an experiment described below (Kim, Lee, et al., 2024; Kim, Oh, et al., 2024). The analysis enabled us to identify associations between individual characteristics and cognitive and attitudinal outcomes in situations with discriminatory AI outcomes that are consistent across different identity dimensions of discrimination (between-subject), different discrimination targets (between-subject), and different contexts of AI use (within-subject).

Participants of the experiment were randomly assigned to one of the four experimental conditions: 2 (identity dimensions: race vs. gender) × 2 (discrimination target: self vs. other). During the experiment, participants were shown the same set of nine scenarios, but the identity dimension and the discrimination target varied depending on the condition. First, regarding the identity dimension of discrimination, participants in the gender condition were asked to think about a friend of a different gender but with similar characteristics (such as race, age, education level, and economic status), while those in the race condition were instructed to consider a friend of a different race but with similar characteristics. Participants were also instructed to keep the friend they chose in mind while considering the scenarios that followed. To verify that participants followed the instruction, they were also asked about the gender of the friend they chose. In terms of the target of discrimination, the discriminatory scenarios of the subject-targeting condition disadvantaged the subject, whereas the scenarios presented in the other-targeting condition disadvantaged the friend. The list of scenarios is presented in Table S1 of the online supporting material.

Each scenario depicted a realistic situation where both a participant and the friend he/she chose were using the same technology based on AI, such as a rating system evaluating job interview performance, but experienced bias in its outcome. The scenarios presented to participants covered a range of topics including media, finance, public service, the labor market, and health and safety. Some of the scenarios were adopted from previous studies, and others were newly created based on real-world incidents reported in news and report (Acikgoz et al., 2020; Binns et al., 2018; A. P. Miller & Hosanagar, 2019; Parra et al., 2021). All scenarios were presented in the same narrative format tested by Parra et al. (2021). Each participant was shown one scenario at a time, and the order of presentation was randomly determined for each participant. The goal of this design was to capture the real-world experiences and exposures that people encounter in their interactions with AI technologies which are often characterized by instances of discrimination and bias (Kim, Lee, et al., 2024; Kim, Oh, et al., 2024).

After viewing each scenario, participants evaluated the outcome presented in it from multiple aspects: fairness of the outcome, trust in the technology generating the outcome, emotion about the outcome, skepticism toward the outcome, and pervasiveness of the outcome in the real world. After viewing all nine scenarios, participants were required to respond to a series of questions gauging their general attitudes toward AI. The questionnaires to measure personal differences and demographic factors were also presented after all scenarios.

The deliberate placement of questionnaires capturing general attitudes following the exposure to the discriminatory scenarios and the assessment of perceptions of discriminatory outcomes was designed with the intention to capture general attitudes toward AI in contexts involving discriminatory outcomes. Essentially, the scenarios and the measures for outcome perception served as an exposure to the discriminatory context for the participants, which enabled them to respond to the questions measuring general attitudes while considering the discriminatory AI situations.

The average time for the completion of the entire experiment was 13.13 minutes ($SD$ = 6.32 minutes). Financial compensation was provided to participants for their participation. The study received an exemption from the Institutional Review Board at the University of California Davis, and participants were required to give their informed consent on the first page of the experiment website. Data collection for the experiment was completed in August 2022.

**Measures**

Outcome fairness, trust, and negative emotion were measured after each scenario and averaged across scenarios for each participant. The summary statistics are as follows: outcome fairness ($M$ = 3.14, $SD$ = 1.09), outcome trust ($M$ = 2.73, $SD$ = 0.94), and negative emotion ($M$ = 4.52, $SD$ = 1.32). All items measuring these variables were adapted from previous work (Araujo et al., 2020; Lee, 2018), which ranged from 1 (lowest) to 7 (highest).

To gauge participants' overall awareness and perception of AI discrimination, two new measures were introduced in this study. First, outcome skepticism reflects the overall tendency of participants to question an AI outcome and cast doubt against it (Kim, Lee, et al., 2024; Kim, Oh, et al., 2024). It was assessed with two items on a 7-point scale from 1 to 7, and participants were asked to rate their agreement with each of the following statements "This outcome is problematic" and "This outcome is questionable." These responses were then averaged to create a single variable. Second, outcome pervasiveness captured the subjective evaluation of the possibility of a certain outcome in everyday life (Kim, Lee, et al., 2024; Kim, Oh, et al., 2024). This variable was assessed on a 7-point scale from 1 (highly unlikely) to 7 (highly likely) using the question "How likely is this outcome to happen in your everyday life?" Both variables were measured after each scenario and averaged across scenarios for each participant: outcome skepticism ($M$ = 5.02, $SD$ = 1.07) and outcome pervasiveness ($M$ = 3.98, $SD$ = 1.19).

Three variables indicating general attitudes toward AI (usefulness, risk, and appreciation) were measured after finishing all scenarios. First, AI risk ($M$ = 4.33, $SD$ = 1.37) is the average response to five items, including "Using AI technologies is risky" and "AI technologies can lead to bad results" (Cox & Cox, 2001). Second, AI usefulness ($M$ = 4.55, $SD$ = 1.46) was accessed with three items, including "Using AI technologies makes me save time" (Nysveen, 2005). Lastly, to measure AI appreciation ($M$ = 2.24, $SD$ = 1.42), a prompt describing a medical situation was adopted from Bigman et al. (2020). The prompt started with "Imagine you are feeling severe shortness of breath and need to go to a hospital. There are two nearby hospitals. You know that both hospitals are running low on supplies and need to prioritize patients…" After reading the prompt, participants were asked about a hospital they would want to go to and reported their preference on a seven-point scale ranging between 1 ("the hospital where the human doctor makes triage decisions") and 7 ("the hospital where an AI-based algorithm makes triage decisions").

Four variables indicating personal differences were also measured after all scenarios. First, for digital self-efficacy ($M$ = 3.18, $SD$ = 1.55), three items from Boerman et al. (2021) were adopted and modified to measure self-efficacy of digital technology use, such as "I am able to protect my personal information when I use digital devices or services." Second, technical knowledge ($M$ = 3.05, $SD$ = 1.53) was measured with three items (Lee & Baykal, 2017). Third, following the previous approach of Araujo et al. (2020), belief in equality ($M$ = 4.38, $SD$ = 1.55) was measured with three items based on the World Value Survey (Ingelhart et al., 2020). Lastly, participants reported their political ideology ($M = 3.64$, $SD = 1.11$) on a standard five-point scale ranging from 1 (very conservative) to 5 (very liberal).

The participants provided information about their education level, which ranged from 1 (less than high school) to 8 (postgraduate or professional degree), and their income level, which ranged from 1 (less than $10,000) to 9 (more than $150,000).

## Statistical Analyses

The hypotheses and research questions were tested using several multivariate linear regressions. In all regression analyses, the identity dimension, discrimination target, and participants' gender and race were controlled. These controls were applied to ensure that the estimated associations between individual characteristics and AI perceptions are more generalizable by accounting for different identity dimensions, discrimination targets, and the social identities of the participants in their estimations. R, an open-source statistical software (version 4.0.3), was used to execute all statistical analyses reported in this study.

On average, the participants responded negatively about the fairness of the outcomes (outcome fairness: $M = 3.14$, $SD = 1.09$) and exhibited overall distrust (outcome trust: $M = 2.73$, $SD = 0.94$) and skepticism (outcome skepticism: $M = 5.02$, $SD = 1.07$) about the outcomes. The participants also expressed negative emotions about the outcomes ($M = 4.52$, $SD = 1.32$). The descriptive statistics of the variables are presented in Table 1.

## Perceptions of Discriminatory AI Outcomes

The associations between individual characteristics and five dependent variables representing outcome perceptions were examined with five separate regression models presented in Table 2 and visualized in Figure 1.

**Outcome Fairness**. Belief in equality was negatively associated with outcome fairness, while income was positively associated with it, adjusting for influences of other characteristics of individuals. The negative association between belief in equality and outcome fairness ($B$ = -0.072, $SE$ = 0.027, $p$ = .007) indicates that as equality belief increased, outcome fairness decreased. Income was positively associated with outcome fairness ($B$ = 0.034, $SE$ = 0.014, $p$ = .016), indicating that the perceived fairness of biased outcomes increased as income level increased.

**Outcome Trust**. Digital self-efficacy, political ideology, and income were positively associated with outcome trust, controlling for other characteristics of individuals. First, digital self-efficacy was positively associated with outcome trust ($B$ = 0.083, $SE$ = 0.018, $p$ < .001). It shows that participants with greater self-efficacy considered that AI made more trustworthy outcomes. Second, the positive association between political ideology and outcome trust ($B$ = 0.088, $SE$ = 0.033, $p$ = .007) shows that participants with more conservative ideologies trusted AI outcomes more. Lastly, income was positively associated with outcome trust ($B$ = 0.036, $SE$ = 0.012, $p$ = .003): as income increased, outcome trust grew.

**Negative Emotion**. Political ideology was negatively associated with negative emotion ($B$ = -0.104, $SE$ = 0.046, $p$ = .023), controlling for other characteristics of individuals. It indicates that more conservative participants exhibited less negative emotion.

**Outcome Skepticism**. Political ideology was negatively associated with outcome skepticism ($B$ = -0.081, $SE$ = 0.037, $p$ = .030), adjusting for the influence of individuals' other characteristics. It indicates that more liberal participants viewed discriminatory outcomes as more problematic.

**Outcome Pervasiveness**. Outcome pervasiveness had positive associations with technical knowledge and age. First, technical knowledge was positively associated with outcome pervasiveness ($B = 0.082$, $SE = 0.023$, $p < .001$), showing that people with more technical knowledge perceived that discriminatory outcomes were more likely to happen in their everyday lives. Also, age was positively associated with outcome pervasiveness ($B = 0.010$, $SE = 0.002$, $p < .001$), indicating that the perceived likelihood of biased outcomes increased as age increases.

## General Attitudes Toward AI Use

The associations between individual characteristics and three dependent variables measuring general perceptions of AI were examined with three separate statistical models presented in Table 3 and visualized in Figure 2.

**AI Usefulness**. Digital self-efficacy, technical knowledge, and education were positively associated with AI usefulness. The positive association between digital self-efficacy and AI usefulness ($B = 0.167$, $SE = 0.027$, $p < .001$) indicates that people with greater self-efficacy perceived AI technologies are more useful. The other positive association between technical knowledge and AI usefulness ($B = 0.162$, $SE = 0.028$, $p < .001$) shows that AI technologies were viewed as more useful for participants who had more domain-specific knowledge of AI technologies. Lastly, as education level increased, perceived usefulness of AI also increased, as evidenced by the positive association between the two variables ($B = 0.085$, $SE = 0.028$, $p = .002$).

**AI Risk**. Digital self-efficacy ($B = -0.202$, $SE = 0.025$, $p < .001$) and belief in equality ($B = -0.080$, $SE = 0.034$, $p = .018$) were both negatively associated with AI risk. It indicates that participants with higher self-efficacy viewed AI as less risky, and so did participants with greater belief in equality.

**AI Appreciation**. Technical knowledge was positively associated with AI appreciation, showing that people with greater knowledge prefer AI over human decision-makers. The other positive association between education and knowledge shows that the levels of general knowledge were linked to the levels of domain-specific knowledge on AI. This finding is not only intuitive but also confirming, as it aligns with our expectations and provides reassurance about the quality of the data.

**Post-hoc Power Analysis**

The relatively large sample size ($N = 1,206$) provided high statistical power for most of the analyses reported above. A post-hoc power analysis was conducted using G*Power 3.1 software (Faul et al., 2009) for the multiple regression models with 11 predictor variables, similar to those presented in Tables 2 and 3. The power analysis confirmed that the sample size provided 93.8% power to detect small effects ($f^2 = 0.02$) at $\alpha = .05$. This level of power is well above the commonly recommended 80% threshold (Cohen et al., 1988), indicating that the present study was well-equipped to detect even small effects. The high power also suggests that non-significant findings are more likely to represent true null effects rather than Type II errors due to insufficient statistical power.

## Discussion

The present study explored the associations between individual characteristics and AI perceptions. This section presents the summary of key findings, the discussion of implications, and the limitations of the current research.

First, digital self-efficacy was positively associated with outcome trust, aligned with H1. However, no statistical significance was identified for the associations between digital self-efficacy and other outcome perceptions (outcome fairness, negative emotion, and outcome skepticism), and thus H1 is only partially supported. The association between digital self-efficacy and outcome pervasiveness was also nonsignificant (RQ4). Regarding general attitudes toward AI, digital self-efficacy was positively associated with AI usefulness and negatively with AI risk. Providing evidence for RQ1, people with higher digital self-efficacy perceived AI more positively in terms of its usefulness and potential risk, even when considering possible discrimination by automated systems. This finding implies that promoting digital literacy skills and enhancing digital self-efficacy could potentially help to maintain or avoid losing people's trust in AI outcomes and their belief in AI usefulness and safety despite the instances of incomplete algorithms and their unfair and biased results in the real world.

The positive association between digital self-efficacy and outcome trust, without statistically significant associations with fairness, negative emotion, or skepticism, may be understood based on the theory of self-efficacy. According to the theory (Bandura, 1977), individuals with higher self-efficacy tend to exhibit greater confidence in their ability to engage effectively with technology. This may explain why individuals with higher digital self-efficacy are more likely to display higher levels of trust in AI outcomes: Their confidence in managing the system may enhance outcome trust. This interpretation also aligns with other results showing that self-efficacy was associated with a reduced perception of risk and an increased perception of usefulness.

However, the associations were not statistically significant between self-efficacy and outcome fairness, negative emotion, and outcome skepticism. While extremely subtle effects might have gone undetected despite the robust sample size, the post-hoc power analysis suggests that the nonsignificant associations are more likely to represent true null findings or effects too small for practical significance, rather than limitations in statistical power. Thus, one can contemplate the theoretical implications of these relationships by considering the role of self-efficacy and its connections to other concepts. Digital self-efficacy does not necessarily translate to higher perceptions of outcome fairness or lower outcome skepticism, as fairness is often perceived as an inherent property of the system rather than one that can be influenced by user ability (Sundar & Kim, 2019). Users with high digital self-efficacy may still recognize and identify unfairness or biases in AI outcomes but feel more capable of navigating or mitigating these issues through their own knowledge and skills. Hence, while they may be more confident in their ability to use the AI system, they can simultaneously hold reservations about its fairness. Similarly, the nonsignificant relationships between digital self-efficacy and negative emotion, and between self-efficacy and skepticism, may result from the fact that users' self-efficacy does not necessarily suppress their negative emotion and skepticism provoked by discriminatory AI outcomes.

Second, regarding technical knowledge, the analysis did not identify statistically significant associations between technical knowledge and four outcome perceptions (outcome fairness, outcome trust, negative emotion, and outcome skepticism), providing no support for H2. However, there was a significant association between technical knowledge and outcome pervasiveness (RQ4): people with more technical knowledge conceive that discriminatory outcomes are likely. This finding can be interpreted as indicating that domain-specific knowledge encourages critical examinations of AI bias in the real world. Lastly, regarding general attitudes, technical knowledge was associated with greater AI usefulness and appreciation (RQ1), suggesting that efforts to educate the public about AI technologies and their potential impacts on society could contribute to shaping positive attitudes toward AI despite exposure to biased AI outcomes.

Third, belief in equality was negatively associated with outcome fairness, but the other three outcome perceptions did not have significant associations with it, providing only partial support for H3. The association between belief in equality and outcome fairness can be interpreted to suggest that individuals with heightened expectations of equality might perceive discriminatory outcomes as less fair, thereby indicating a greater perceived need for equality in the discriminatory situations. The association between technical knowledge and outcome pervasiveness was nonsignificant (RQ4). Regarding the relationship between equality belief and general attitudes (RQ1), there was a negative association between equality belief and AI risk. This association implies that people with a strong desire for equality may still view AI technologies as a way to produce less risky decisions, despite the exposure to discriminatory situations. Future research should further discover and validate the mechanisms behind this relationship.

Fourth, lending considerable support to H4, more liberal ideologies were associated with lower outcome trust, higher negative emotion, and greater skepticism. Although statistical significance was not identified, the sign of the association between outcome fairness and ideology was also aligned with the hypothesis. However, statistical significance was not detected for the associations between political ideology and outcome pervasiveness (RQ4). Statistical significance was also not identified for the associations between the three measures of AI attitudes and political ideology (RQ1).

Evidence for RQ1 presented above generally aligns with a previous study conducted in a European country (Araujo et al., 2020), suggesting that similar mechanisms may operate across different national and cultural contexts. For example, consistent with the previous work, digital self-efficacy was positively associated with AI usefulness and negatively with AI risk, and technical knowledge was associated with greater AI usefulness and appreciation. Confirming the past finding, this study also found a negative association between equality belief and AI risk.

Fifth, this study provides evidence regarding the role of demographic factors in shaping the perceptions of discriminatory outcomes (RQ2). The positive associations between income, and perceived trust and fairness; and between age and outcome pervasiveness not only align with previous findings on the perceptions of new technologies (Coughlin et al., 2007; McComas & Besley, 2011) but also imply that economic inequalities and generational gaps may be closely connected to the cognitive gaps in understanding problematic outcomes produced by the technologies. It is noteworthy that these associations were still statistically significant even when the two plausible confounding factors, education and technological knowledge, were controlled for.

Sixth, regarding the relationships between demographic factors and general attitudes toward AI (RQ3), education appeared to be a key factor connected with both AI usefulness and AI appreciation. Specifically, people with higher education levels viewed AI as more useful and were more likely to prefer AI-based decisions over human decisions.

The present study is not without limitations. Firstly, the sample of participants was not representative of the entire U.S. population, which means that the findings may not be applicable to the general population. This is a common issue for studies based on convenience samples, and further research with a more diverse and representative sample would be necessary to confirm these results. Second, the study examined perceptions of discriminatory outcomes in a hypothetical scenario. Future research should measure perceptions and reactions to algorithm outcomes in more realistic and interactive settings. Third, the scenarios in this study were focused on consumer and user contexts, but it would be useful to conduct further research to examine potential algorithmic bias in other areas, such as the justice system. Fourth, the measures for outcome skepticism and outcome pervasiveness relied on a small number of items and required further investigation of their validity. Hence, the results concerning these outcomes should be interpreted with caution. Lastly, this study is a secondary analysis and was not pre-registered, highlighting the need for future studies to follow rigorous open-science practices and potentially conduct pilot tests to refine the research design and ensure the validity of the measures used.

**Design Implications for AI Systems**

Based on the present research's findings, several design implications can be drawn for practitioners and designers aiming to improve the experiences of AI system users. First, because the findings indicate that users with higher digital self-efficacy trust AI outcomes more and perceive higher levels of usefulness and lower levels of risk from AI, even when facing potential biases or unfair outcomes, enhancing digital self-efficacy could improve user experiences. Implementing transparency features could be helpful in this sense by increasing self-efficacy, such as clear explanations of how AI decisions are made, what data is used, and how users can contest or adjust AI outcomes (Binns et al., 2018; Veale et al., 2018). Embedding interactive tutorials or feedback mechanisms may also enable users to navigate AI systems better, improving their digital self-efficacy and trust (Liu et al., 2021). In line with this design implication, Wang et al. (2020) emphasized procedural transparency's role in shaping fairness perceptions. Angerschmid et al. (2022) also reported that providing clear explanations of AI outcomes increases trust and fairness perceptions. Similarly, Leichtmann et al. (2023) showed that users who received explanations of AI predictions were better able to understand and trust the system.

Second, fostering digital literacy and enhancing technological knowledge can enable users to effectively engage with AI systems, as implied by the connection between technical knowledge and the perception of discriminatory outcomes. The current research has shown that technical knowledge was strongly associated with the perception of discriminatory outcomes as pervasiveness, suggesting that users with higher technical knowledge are more adept at recognizing AI bias. It was also connected with higher perceived AI usefulness and appreciation. Aligned with past research indicating that an enhanced understanding of AI systems leads to more critical engagement with technology (Araujo et al., 2020; Lee & Baykal, 2017), these findings imply that integrating learning modules that help build technical knowledge may allow users to better identify potential biases while maintaining positive attitudes toward AI systems. Personalized learning experiences that provide tailored step-by-step guidance and contextual information may facilitate this process (Wang et al., 2020). These approaches may not only demystify AI but also empower users to make informed decisions and promote more critical examinations of AI applications.

Third, the associations between belief in equality and political ideology with AI perceptions offer further design implications. Users with strong beliefs in equality and more liberal political ideologies are particularly sensitive to discriminatory outcomes, perceiving these outcomes as less fair and trustworthy and responding with greater skepticism and negative emotion. Thus, designers may incorporate adaptive features that actively demonstrate fairness and accountability, including mechanisms allowing users of diverse beliefs to report biases and view adjustments made to AI decisions (Schwartz et al., 2022). Such features will not only serve users' desire for fairness but also engender trust by showing that the system's operations align with ethical standards. Furthermore, tools simulating and presenting different decision-making scenarios could illustrate how the AI arrives at its conclusions under varying conditions, helping users understand the rationale behind the decisions and enhancing transparency.

Lastly, demographic factors such as income, age, and education play a crucial role in shaping user perceptions of AI systems and thus have important design implications. The current findings indicate that individuals with lower income levels perceive discriminatory AI outcomes as more untrustworthy and unfair, while older users view such outcomes as more pervasive. Lower education levels are also linked to reduced perceptions of AI usefulness and appreciation. These results highlight the need for designers to consider how economic, educational, and generational disparities influence cognitive engagement with AI, echoing prior research on this issue (Kieslich et al., 2022; Sin et al., 2021; Wolf & Ringland, 2020). To bridge these gaps, AI systems could incorporate flexible design features that accommodate users across different demographic backgrounds, especially those from underserved, marginalized, or vulnerable groups. For example, AI systems might offer tiered explanations of decision-making processes, ranging from basic, accessible overviews suitable for users less familiar with or knowledgeable about AI technologies, to more complex technical breakdowns. By offering multi-level engagement tools that adapt to various income, age, and education levels, designers can create AI experiences that are equitable, socially aware, contextualized, and tailored to the diverse needs of the user base.

## Conclusion

The findings of this study carry important implications. Promoting digital literacy skills and enhancing digital self-efficacy could potentially help maintain trust in AI outcomes and beliefs in AI usefulness and safety or prevent losing them, despite instances of flawed AI algorithms and their unfair and biased results in the real world. Efforts to educate the public about AI technologies and their potential impacts on society could also help shape positive attitudes toward AI. Furthermore, this study underscores the importance of recognizing and addressing the potential for economic inequalities and generational gaps in understanding problematic outcomes produced by technology. Also, the findings imply the potential difference between cognitive mechanisms governing perceptions of biased AI outcomes and those involving general attitudes toward the technology.

Overall, this study represents one of the initial attempts to unpack the socio-technological system, where complex interactions occur between social hierarchies, divisions, and machines that reflect and worsen the disparities. Specifically, it demonstrates that theorizing the emerging questions about perceptions of biased and unequal AI on the basis of long-standing social scientific investigations on intergroup interactions, attribution processes, and cognitive biases can be fruitful in this new research frontier.

## Conflict of Interest

There is no conflict of interest regarding the research presented in this paper.

## Data Availability

The datasets generated during and/or analyzed during the current study are available in the OSF repository, https://osf.io/bph82/?view_only=57da7439b0bd42159a93353a39220e4c

# Reference

Acikgoz, Y., Davison, K. H., Compagnone, M., & Laske, M. (2020). Justice perceptions of artificial intelligence in selection. *International Journal of Selection and Assessment*, *28*(4), 399–416. https://doi.org/10.1111/ijsa.12306

Aesaert, K., & van Braak, J. (2014). Exploring factors related to primary school pupils' ICT self-efficacy: A multilevel approach. *Computers in Human Behavior*, *41*, 327–341. https://doi.org/10.1016/j.chb.2014.10.006

Angerschmid, A., Zhou, J., Theuermann, K., Chen, F., & Holzinger, A. (2022). Fairness and Explanation in AI-Informed Decision Making. *Machine Learning and Knowledge Extraction*, *4*(2), 556–579. https://doi.org/10.3390/make4020026

Araujo, T., Helberger, N., Kruikemeier, S., & de Vreese, C. H. (2020). In AI we trust? Perceptions about automated decision-making by artificial intelligence. *AI & SOCIETY*, *35*(3), 611–623. https://doi.org/10.1007/s00146-019-00931-w

Asplund, J., Eslami, M., Sundaram, H., Sandvig, C., & Karahalios, K. (2020). Auditing Race and Gender Discrimination in Online Housing Markets. *Proceedings of the Fourteenth International AAAI Conference on Web and Social Media*, 24–35.

Bagci, S. C., Çelebi, E., & Karaköse, S. (2017). Discrimination Towards Ethnic Minorities: How Does it Relate to Majority Group Members' Outgroup Attitudes and Support for Multiculturalism. *Social Justice Research*, *30*(1), 1–22. https://doi.org/10.1007/s11211-017-0281-6

Bandura, A. (1977). Self-efficacy: Toward a unifying theory of behavioral change. *Psychological Review*, *84*(2), 191.

Barocas, S., Hardt, M., & Narayanan, A. (2019). *Fairness and machine learning*.
https://fairmlbook.org

Benjamin, R. (2019). *Race after technology: Abolitionist tools for the new Jim code*. Polity.

Bigman, Y., Gray, K., Waytz, A., Arnestad, M., & Wilson, D. (2020). *Algorithmic Discrimination Causes Less Moral Outrage than Human Discrimination* [Preprint].
PsyArXiv. https://doi.org/10.31234/osf.io/m3nrp

Binns, R., Van Kleek, M., Veale, M., Lyngs, U., Zhao, J., & Shadbolt, N. (2018, January 31).
*"It's Reducing a Human Being to a Percentage"; Perceptions of Justice in Algorithmic Decisions*. ACM CHI Conference on Human Factors in Computing Systems.
https://doi.org/10.31235/osf.io/9wqxr

Boerman, S. C., Kruikemeier, S., & Zuiderveen Borgesius, F. J. (2021). Exploring Motivations for Online Privacy Protection Behavior: Insights From Panel Data. *Communication Research*, *48*(7), 953–977. https://doi.org/10.1177/0093650218800915

Castelo, N., & Ward, A. F. (2021). Conservatism predicts aversion to consequential Artificial Intelligence. *PLOS ONE*, *16*(12), e0261467.
https://doi.org/10.1371/journal.pone.0261467

Chen, L., Ma, R., Hannák, A., & Wilson, C. (2018). Investigating the Impact of Gender on Rank in Resume Search Engines. *Proceedings of the 2018 CHI Conference on Human Factors in Computing Systems*, 1–14. https://doi.org/10.1145/3173574.3174225

Choung, H., David, P., & Ross, A. (2023). Trust in AI and Its Role in the Acceptance of AI Technologies. *International Journal of Human–Computer Interaction*, *39*(9), 1727–1739.
https://doi.org/10.1080/10447318.2022.2050543

Choung, H., Seberger, J. S., & David, P. (2023). When AI is Perceived to Be Fairer than a Human: Understanding Perceptions of Algorithmic Decisions in a Job Application Context. *International Journal of Human–Computer Interaction*, *0*(0), 1–18. https://doi.org/10.1080/10447318.2023.2266244

Cohen, J., Mutz, D., Price, V., & Gunther, A. (1988). Perceived Impact of Defamation: An Experiment on Third-Person Effects. *Public Opinion Quarterly*, *52*(2), 161. https://doi.org/10.1086/269092

Coughlin, J. F., D'Ambrosio, L. A., Reimer, B., & Pratt, M. R. (2007). Older Adult Perceptions of Smart Home Technologies: Implications for Research, Policy & Market Innovations in Healthcare. *2007 29th Annual International Conference of the IEEE Engineering in Medicine and Biology Society*, 1810–1815. https://doi.org/10.1109/IEMBS.2007.4352665

Cox, D., & Cox, A. D. (2001). Communicating the Consequences of Early Detection: The Role of Evidence and Framing. *Journal of Marketing*, *65*(3), 91–103. https://doi.org/10.1509/jmkg.65.3.91.18336

Diakopoulos, N., & Koliska, M. (2017). Algorithmic Transparency in the News Media. *Digital Journalism*, *5*(7), 809–828. https://doi.org/10.1080/21670811.2016.1208053

Dietvorst, B. J., Simmons, J. P., & Massey, C. (2014). Algorithm aversion: People erroneously avoid algorithms after seeing them err. *Journal of Experimental Psychology: General*, *144*(1), 114. https://doi.org/10.1037/xge0000033

Dolata, M., Feuerriegel, S., & Schwabe, G. (2022). A sociotechnical view of algorithmic fairness. *Information Systems Journal*, *32*(4), 754–818. https://doi.org/10.1111/isj.12370

Faul, F., Erdfelder, E., Buchner, A., & Lang, A.-G. (2009). Statistical power analyses using G*Power 3.1: Tests for correlation and regression analyses. *Behavior Research Methods*, *41*(4), 1149–1160. https://doi.org/10.3758/BRM.41.4.1149

Gong, F., Xu, J., & Takeuchi, D. T. (2017). Racial and Ethnic Differences in Perceptions of Everyday Discrimination. *Sociology of Race and Ethnicity*, *3*(4), 506–521. https://doi.org/10.1177/2332649216681587

Hajigholam Saryazdi, A. (2024). Algorithm Bias and Perceived Fairness: A Comprehensive Scoping Review. *Proceedings of the 2024 Computers and People Research Conference*, 1–9. https://doi.org/10.1145/3632634.3655848

Helberger, N., Araujo, T., & de Vreese, C. H. (2020). Who is the fairest of them all? Public attitudes and expectations regarding automated decision-making. *Computer Law & Security Review*, *39*, 105456. https://doi.org/10.1016/j.clsr.2020.105456

Hooker, S. (2021). Moving beyond "algorithmic bias is a data problem." *Patterns*, *2*(4), 100241. https://doi.org/10.1016/j.patter.2021.100241

Imana, B., Korolova, A., & Heidemann, J. (2021). Auditing for Discrimination in Algorithms Delivering Job Ads. *Proceedings of the Web Conference 2021*, 3767–3778. https://doi.org/10.1145/3442381.3450077

Ingelhart, R., Haerpfer, C. W., Moreno, A., Welzel, C., Kizilova, K., Diez-Medrano, J., Lagos, M., Norris, P., Ponarin, E., & Puranen, B. (2020). *World Values Survey Wave 5 (2005-2009)* (Version 20180912) [Dataset]. World Values Survey Association. https://doi.org/10.14281/18241.7

Jha, S., & Topol, E. J. (2016). Adapting to Artificial Intelligence: Radiologists and Pathologists as Information Specialists. *JAMA*, *316*(22), 2353–2354. https://doi.org/10.1001/jama.2016.17438

Jost, J. T., Federico, C. M., & Napier, J. L. (2009). Political Ideology: Its Structure, Functions, and Elective Affinities. *Annual Review of Psychology*, *60*(1), 307–337. https://doi.org/10.1146/annurev.psych.60.110707.163600

Kennedy, L. W., Caplan, J. M., & Piza, E. (2011). Risk Clusters, Hotspots, and Spatial Intelligence: Risk Terrain Modeling as an Algorithm for Police Resource Allocation Strategies. *Journal of Quantitative Criminology*, *27*(3), 339–362. https://doi.org/10.1007/s10940-010-9126-2

Kieslich, K., Keller, B., & Starke, C. (2022). Artificial intelligence ethics by design. Evaluating public perception on the importance of ethical design principles of artificial intelligence. *Big Data & Society*, *9*(1), 20539517221092956. https://doi.org/10.1177/20539517221092956

Kim, S., Lee, J., & Oh, P. (2024). Questioning artificial intelligence: How racial identity shapes the perceptions of algorithmic bias. *International Journal of Communication*, *18*, 677–699.

Kim, S., Oh, P., & Lee, J. (2024). Algorithmic gender bias: Investigating perceptions of discrimination in automated decision-making. *Behaviour & Information Technology*, 1–14. https://doi.org/10.1080/0144929X.2024.2306484

Kissinger, H., Schmidt, E., Huttenlocher, D. P., & Schouten, S. (2021). *The age of AI: And our human future* (First edition). Little Brown and Company.

Koenecke, A., Nam, A., Lake, E., Nudell, J., Quartey, M., Mengesha, Z., Toups, C., Rickford, J. R., Jurafsky, D., & Goel, S. (2020). Racial disparities in automated speech recognition. *Proceedings of the National Academy of Sciences*, *117*(14), 7684–7689. https://doi.org/10.1073/pnas.1915768117

Lee, M. K. (2018). Understanding perception of algorithmic decisions: Fairness, trust, and emotion in response to algorithmic management. *Big Data & Society*, *5*(1), 2053951718756684. https://doi.org/10.1177/2053951718756684

Lee, M. K., & Baykal, S. (2017). Algorithmic Mediation in Group Decisions: Fairness Perceptions of Algorithmically Mediated vs. Discussion-Based Social Division. *Proceedings of the 2017 ACM Conference on Computer Supported Cooperative Work and Social Computing*, 1035–1048. https://doi.org/10.1145/2998181.2998230

Lee, M. K., & Rich, K. (2021). Who Is Included in Human Perceptions of AI?: Trust and Perceived Fairness around Healthcare AI and Cultural Mistrust. *Proceedings of the 2021 CHI Conference on Human Factors in Computing Systems*, 1–14. https://doi.org/10.1145/3411764.3445570

Leichtmann, B., Humer, C., Hinterreiter, A., Streit, M., & Mara, M. (2023). Effects of Explainable Artificial Intelligence on trust and human behavior in a high-risk decision task. *Computers in Human Behavior*, *139*, 107539. https://doi.org/10.1016/j.chb.2022.107539

Liu, H., Lai, V., & Tan, C. (2021). Understanding the Effect of Out-of-distribution Examples and Interactive Explanations on Human-AI Decision Making. *Proc. ACM Hum.-Comput. Interact.*, *5*(CSCW2), 408:1-408:45. https://doi.org/10.1145/3479552

Logg, J. M., Minson, J. A., & Moore, D. A. (2019). Algorithm appreciation: People prefer algorithmic to human judgment. *Organizational Behavior and Human Decision Processes*, *151*, 90–103. https://doi.org/10.1016/j.obhdp.2018.12.005

Longoni, C., Bonezzi, A., & Morewedge, C. K. (2019). Resistance to Medical Artificial Intelligence. *Journal of Consumer Research*, *46*(4), 629–650. https://doi.org/10.1093/jcr/ucz013

Mack, E. A., Miller, S. R., Chang, C.-H., Van Fossen, J. A., Cotten, S. R., Savolainen, P. T., & Mann, J. (2021). The politics of new driving technologies: Political ideology and autonomous vehicle adoption. *Telematics and Informatics*, *61*, 101604. https://doi.org/10.1016/j.tele.2021.101604

Makhortykh, M., Urman, A., & Ulloa, R. (2021). Detecting Race and Gender Bias in Visual Representation of AI on Web Search Engines. In L. Boratto, S. Faralli, M. Marras, & G. Stilo (Eds.), *Advances in Bias and Fairness in Information Retrieval* (pp. 36–50). Springer International Publishing. https://doi.org/10.1007/978-3-030-78818-6_5

McComas, K. A., & Besley, J. C. (2011). Fairness and Nanotechnology Concern. *Risk Analysis*, *31*(11), 1749–1761. https://doi.org/10.1111/j.1539-6924.2011.01676.x

Menezes, H. F., Ferreira, A. S. C., Pereira, E. T., & Gomes, H. M. (2021). Bias and Fairness in Face Detection. *2021 34th SIBGRAPI Conference on Graphics, Patterns and Images (SIBGRAPI)*, 247–254. https://doi.org/10.1109/SIBGRAPI54419.2021.00041

Miller, A. P., & Hosanagar, K. (2019, November 8). How Targeted Ads and Dynamic Pricing Can Perpetuate Bias. *Harvard Business Review*. https://hbr.org/2019/11/how-targeted-ads-and-dynamic-pricing-can-perpetuate-bias

Miller, J. (2020, September 18). Is an Algorithm Less Racist Than a Loan Officer? *The New York Times*. https://www.nytimes.com/2020/09/18/business/digital-mortgages.html

Nissan, E. (2017). Digital technologies and artificial intelligence's present and foreseeable impact on lawyering, judging, policing and law enforcement. *AI & SOCIETY*, *32*(3), 441–464. https://doi.org/10.1007/s00146-015-0596-5

Noble, S. U. (2018). *Algorithms of oppression: How search engines reinforce racism*. New York University Press.

Nysveen, H. (2005). Intentions to Use Mobile Services: Antecedents and Cross-Service Comparisons. *Journal of the Academy of Marketing Science*, *33*(3), 330–346. https://doi.org/10.1177/0092070305276149

Obermeyer, Z., Powers, B., Vogeli, C., & Mullainathan, S. (2019). Dissecting racial bias in an algorithm used to manage the health of populations. *Science*, *366*(6464), 447–453. https://doi.org/10.1126/science.aax2342

Ochmann, J., Michels, L., Tiefenbeck, V., Maier, C., & Laumer, S. (2024). Perceived algorithmic fairness: An empirical study of transparency and anthropomorphism in algorithmic recruiting. *Information Systems Journal*, *34*(2), 384–414. https://doi.org/10.1111/isj.12482

O'Neil, C. (2016). *Weapons of math destruction: How big data increases inequality and threatens democracy* (First edition). Crown.

Parra, C. M., Gupta, M., & Dennehy, D. (2021). Likelihood of Questioning AI-based Recommendations Due to Perceived Racial/Gender Bias. *IEEE Transactions on Technology and Society*, 1–1. IEEE Transactions on Technology and Society. https://doi.org/10.1109/TTS.2021.3120303

Peng, Y. (2020). The ideological divide in public perceptions of self-driving cars. *Public Understanding of Science*, *29*(4), 436–451. https://doi.org/10.1177/0963662520917339

Rader, E., & Gray, R. (2015). Understanding User Beliefs About Algorithmic Curation in the Facebook News Feed. *Proceedings of the 33rd Annual ACM Conference on Human Factors in Computing Systems*, 173–182. https://doi.org/10.1145/2702123.2702174

Rimal, R. N. (2001). Perceived Risk and Self-Efficacy as Motivators: Understanding Individuals' Long-Term Use of Health Information. *Journal of Communication*, *51*(4), 633–654. https://doi.org/10.1111/j.1460-2466.2001.tb02900.x

Robbennolt, J. K. (2000). Outcome Severity and Judgments of "Responsibility": A Meta-Analytic Review1. *Journal of Applied Social Psychology*, *30*(12), 2575–2609. https://doi.org/10.1111/j.1559-1816.2000.tb02451.x

Said, N., Potinteu, A. E., Brich, I. R., Buder, J., Schumm, H., & Huff, M. (2022). *An Artificial Intelligence Perspective: How Knowledge and Confidence Shape Risk and Opportunity Perception* [Preprint]. PsyArXiv. https://doi.org/10.31234/osf.io/5zvha

Schiff, D. S., Schiff, K. J., & Pierson, P. (2022). Assessing public value failure in government adoption of artificial intelligence. *Public Administration*, *100*(3), 653–673. https://doi.org/10.1111/padm.12742

Schmitt, M. T., Branscombe, N. R., & Postmes, T. (2003). Women's emotional responses to the pervasiveness of gender discrimination. *European Journal of Social Psychology*, *33*(3), 297–312. https://doi.org/10.1002/ejsp.147

Schroeder, D. A., & Linder, D. E. (1976). Effects of actor's causal role, outcome severity, and knowledge of prior accidents upon attributions of responsibility. *Journal of Experimental Social Psychology*, *12*(4), 340–356. https://doi.org/10.1016/S0022-1031(76)80003-0

Schwartz, R., Vassilev, A., Greene, K., Perine, L., Burt, A., & Hall, P. (2022). *Towards a standard for identifying and managing bias in artificial intelligence* (No. NIST SP 1270; p. NIST SP 1270). National Institute of Standards and Technology (U.S.). https://doi.org/10.6028/NIST.SP.1270

Sin, J., L. Franz, R., Munteanu, C., & Barbosa Neves, B. (2021). Digital Design Marginalization: New Perspectives on Designing Inclusive Interfaces. *Proceedings of the 2021 CHI Conference on Human Factors in Computing Systems*, 1–11. https://doi.org/10.1145/3411764.3445180

Skarlicki, D. P., & Folger, R. (1997). Retaliation in the Workplace: The Roles of Distributive, Procedural, and Interactional Justice. *Journal of Applied Psychology*.

Starke, C., Baleis, J., Keller, B., & Marcinkowski, F. (2022). Fairness perceptions of algorithmic decision-making: A systematic review of the empirical literature. *Big Data & Society*, *9*(2), 20539517221115189. https://doi.org/10.1177/20539517221115189

Stouten, J., De Cremer, D., & van Dijk, E. (2006). Violating Equality in Social Dilemmas: Emotional and Retributive Reactions as a Function of Trust, Attribution, and Honesty. *Personality and Social Psychology Bulletin*, *32*(7), 894–906. https://doi.org/10.1177/0146167206287538

Stroebe, K., Dovidio, J. F., Barreto, M., Ellemers, N., & John, M.-S. (2011). Is the world a just place? Countering the negative consequences of pervasive discrimination by affirming the world as just: Negative consequences of discrimination. *British Journal of Social Psychology*, *50*(3), 484–500. https://doi.org/10.1348/014466610X523057

Sundar, S. S. (2008). The MAIN Model: A Heuristic Approach to Understanding Technology Effects on Credibility. In M. J. Metzger & A. J. Flanagin (Eds.), *Digital Media, Youth, and Credibility* (pp. 73–100). The MIT Press.

Sundar, S. S., & Kim, J. (2019). Machine Heuristic: When We Trust Computers More than Humans with Our Personal Information. *Proceedings of the 2019 CHI Conference on Human Factors in Computing Systems*, 1–9. https://doi.org/10.1145/3290605.3300768

Swim, J. K., Scott, E. D., Sechrist, G. B., Campbell, B., & Stangor, C. (2003). The role of intent and harm in judgments of prejudice and discrimination. *Journal of Personality and Social Psychology*, *84*(5), 944–959. https://doi.org/10.1037/0022-3514.84.5.944

Thune, J. (2022). Google may send Republicans to spam, but we're holding Big Tech accountable | Fox News. *Fox News*. https://www.foxnews.com/opinion/google-may-send-republicans-spam-holding-big-tech-accountable

Thurman, N., Moeller, J., Helberger, N., & Trilling, D. (2019). My Friends, Editors, Algorithms, and I: Examining audience attitudes to news selection. *Digital Journalism*, *7*(4), 447–469. https://doi.org/10.1080/21670811.2018.1493936

Veale, M., Van Kleek, M., & Binns, R. (2018). Fairness and Accountability Design Needs for Algorithmic Support in High-Stakes Public Sector Decision-Making. *Proceedings of the 2018 CHI Conference on Human Factors in Computing Systems*, 1–14. https://doi.org/10.1145/3173574.3174014

Verma, P. (2022, July 19). These robots were trained on AI. They became racist and sexist. *Washington Post*. https://www.washingtonpost.com/technology/2022/07/16/racist-robots-ai/

Vlasceanu, M., & Amodio, D. M. (2022). Propagation of societal gender inequality by internet search algorithms. *Proceedings of the National Academy of Sciences*, *119*(29), e2204529119. https://doi.org/10.1073/pnas.2204529119

Walster, E. (1966). Assignment of responsibility for an accident. *Journal of Personality and Social Psychology*, *3*(1), 73–79. https://doi.org/10.1037/h0022733

Wang, R., Harper, F. M., & Zhu, H. (2020). Factors Influencing Perceived Fairness in Algorithmic Decision-Making: Algorithm Outcomes, Development Procedures, and Individual Differences. *Proceedings of the 2020 CHI Conference on Human Factors in Computing Systems*, 1–14. https://doi.org/10.1145/3313831.3376813

Wolf, C. T., & Ringland, K. E. (2020). Designing accessible, explainable AI (XAI) experiences. *ACM SIGACCESS Accessibility and Computing*, *125*, 1–1. https://doi.org/10.1145/3386296.3386302

Yu, K.-H., & Kohane, I. S. (2019). Framing the challenges of artificial intelligence in medicine. *BMJ Quality & Safety*, *28*(3), 238–241. https://doi.org/10.1136/bmjqs-2018-008551

Zhang, B., & Dafoe, A. (2019). *Artificial Intelligence: American Attitudes and Trends* (SSRN Scholarly Paper No. 3312874). https://doi.org/10.2139/ssrn.3312874

Table 1. Summary Statistics of Variables ($N$ = 1,206)

| Outcome perceptions | $M$ ($SD$) | General attitudes | $M$ ($SD$) | Personal differences | $M$ ($SD$) |
|---|---|---|---|---|---|
| Outcome fairness | 3.136 (1.091) | AI usefulness | 4.547 (1.462) | Digital self-efficacy | 3.178 (1.549) |
| Outcome trust | 2.726 (0.945) | AI risk | 4.332 (1.371) | Tech knowledge | 3.045 (1.529) |
| Negative emotion | 4.516 (1.319) | AI appreciation | 3.981 (1.189) | Belief in equality | 4.378 (1.553) |
| Outcome skepticism | 5.023 (1.066) | | | Political ideology | 3.638 (1.108) |
| Outcome pervasiveness | 3.981 (1.189) | | | | |

Table 2. The associations between individual characteristics and the perceptions of discriminatory AI outcomes

| | Outcome fairness | Outcome trust | Negative emotion | Outcome skepticism | Outcome pervasiveness |
|---|---|---|---|---|---|
| *Personal differences* | | | | | |
| Dig. self-efficacy | 0.026 (0.020) | 0.083*** (0.018) | 0.007 (0.025) | -0.021 (0.020) | -0.024 (0.022) |
| Tech knowledge | 0.031 (0.021) | 0.033 (0.018) | 0.002 (0.025) | -0.019 (0.021) | 0.082*** (0.023) |
| Belief in equality | -0.072** (0.027) | -0.025 (0.023) | 0.061 (0.032) | 0.045 (0.027) | 0.0003 (0.030) |
| Ideology (liberal) | -0.070 (0.038) | -0.088** (0.033) | 0.104* (0.046) | 0.081* (0.037) | -0.024 (0.042) |
| *Demographic factors* | | | | | |
| Education | 0.003 (0.021) | 0.017 (0.018) | 0.036 (0.025) | 0.005 (0.021) | 0.0003 (0.023) |
| Age | 0.002 (0.002) | -0.0001 (0.002) | -0.0005 (0.003) | 0.004 (0.002) | 0.010*** (0.002) |
| Income | 0.034* (0.014) | 0.036** (0.012) | -0.031 (0.017) | -0.026 (0.014) | 0.006 (0.016) |
| *Control variables* | | | | | |
| Identity (gender) | 0.067 (0.061) | 0.059 (0.053) | -0.210** (0.074) | -0.059 (0.060) | 0.098 (0.067) |
| Disc. target (self) | -0.313*** (0.061) | -0.117* (0.053) | 0.409*** (0.074) | -0.092 (0.060) | -0.291*** (0.067) |
| Race (white) | 0.090 (0.071) | 0.054 (0.061) | -0.036 (0.086) | -0.105 (0.070) | -0.277*** (0.078) |
| Gender (male) | 0.114 (0.064) | 0.163** (0.055) | -0.313*** (0.078) | -0.229*** (0.064) | -0.092 (0.071) |
| Constant | 2.832*** (0.259) | 1.905*** (0.224) | 4.522*** (0.313) | 5.368*** (0.256) | 3.680*** (0.286) |

*Note.* *** $p < 0.001$, ** $p < 0.01$, * $p < 0.05$. $N = 1,206$. Values are unstandardized coefficients of multivariate linear regression models, and standard errors are reported in parentheses.

Table 3. The associations between individual characteristics and AI perceptions

|  | AI usefulness | AI risk | AI appreciation |
|---|---|---|---|
| *Personal differences* | | | |
| Dig. self-efficacy | 0.167*** (0.027) | -0.202*** (0.025) | 0.029 (0.027) |
| Technical knowledge | 0.162*** (0.028) | -0.024 (0.026) | 0.118*** (0.028) |
| Belief in equality | -0.006 (0.036) | -0.080* (0.034) | 0.067 (0.035) |
| Ideology (liberal) | 0.029 (0.050) | 0.068 (0.047) | 0.003 (0.050) |
| *Demographic factors* | | | |
| Education | 0.085** (0.028) | 0.019 (0.026) | 0.066* (0.028) |
| Age | -0.002 (0.003) | 0.001 (0.003) | -0.005 (0.003) |
| Income | 0.029 (0.019) | -0.026 (0.018) | -0.002 (0.019) |
| *Control variables* | | | |
| Identity (gender) | -0.083 (0.081) | -0.007 (0.076) | -0.114 (0.081) |
| Disc. target (self) | -0.101 (0.081) | -0.053 (0.076) | -0.076 (0.081) |
| Race (white) | -0.063 (0.094) | -0.029 (0.089) | -0.102 (0.094) |
| Gender (male) | 0.097 (0.085) | -0.266*** (0.081) | 0.202* (0.085) |
| Constant | 3.173*** (0.343) | 5.721*** (0.324) | 1.455*** (0.342) |

*Note.* *** $p < 0.001$, ** $p < 0.01$, * $p < 0.05$. $N = 1,206$. Values are unstandardized coefficients of multivariate linear regression models, and standard errors are reported in parentheses.

Figure 1. Individual characteristics and the perceptions of discriminatory AI outcomes

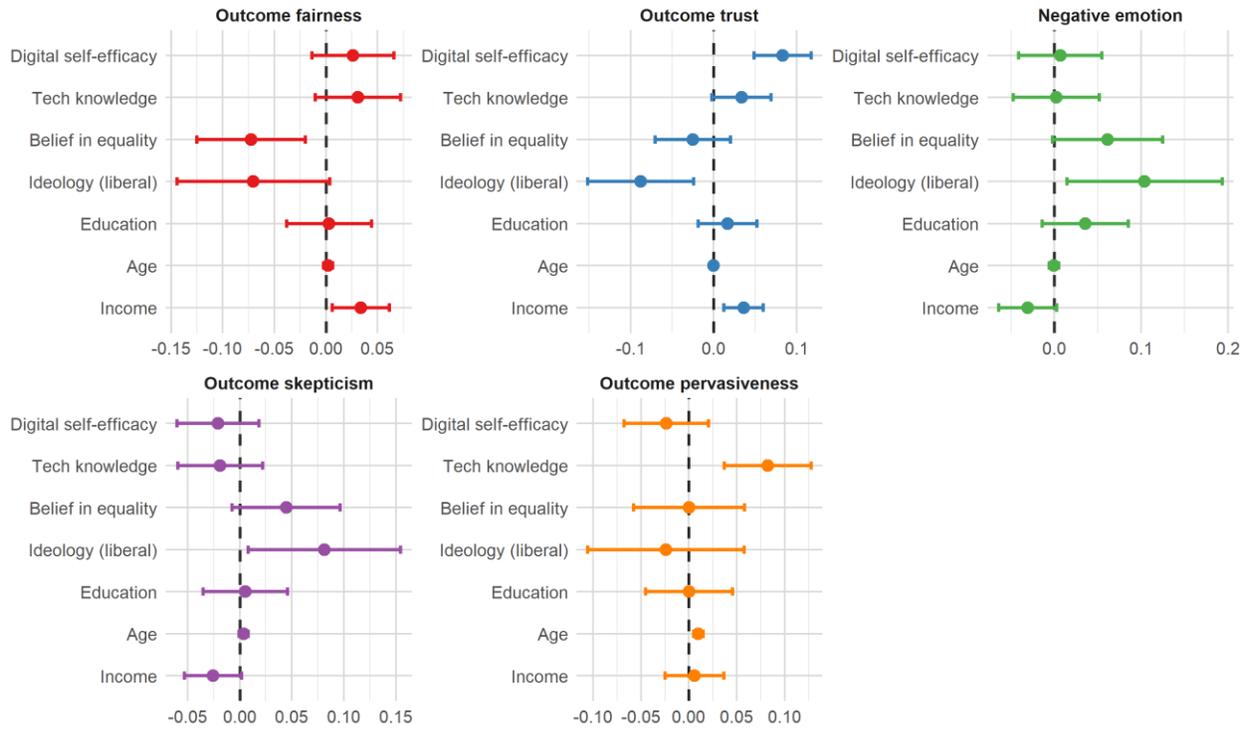

*Note.* Dots indicate coefficients estimated from multivariate linear regression models. Error bars indicate 95% confidence intervals.

Figure 2. Individual characteristics and AI attitudes

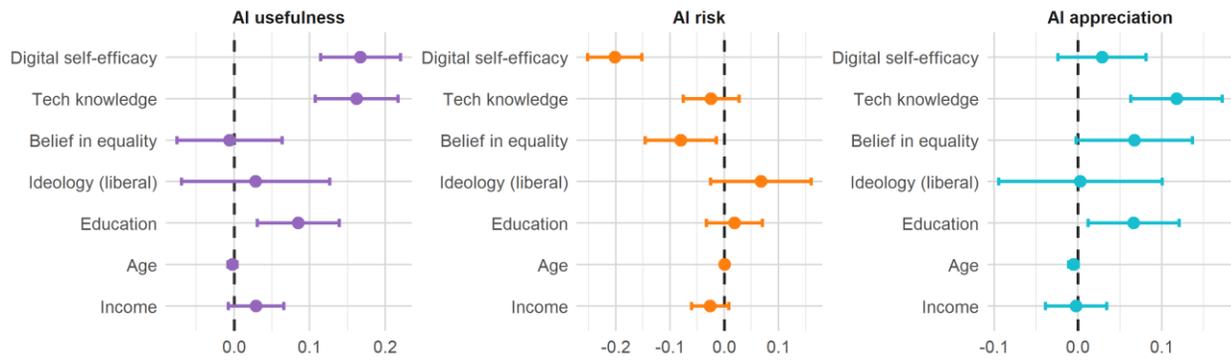

*Note*. Dots indicate coefficients estimated from multivariate linear regression models. Error bars indicate 95% confidence intervals.